\documentclass[11pt,twoside]{article}

\usepackage{asp2006}

\begin{document}
\title{Pathway Toward an Infrared Interferometer}   
\author{Jean Schneider}   
\affil{LUTh - Paris Observatory, Meudon, France}    

\begin{abstract} 
We propose a  realistic pathway to satisfy two goals, thermal infrared studies of Earth-like
exoplanets and interferometric architectures.
\end{abstract}

\section{An Invitable Need for the Long Term: Interferometry.}
After the first low resolution spectro-imaging of exoplanets, a first obvious step will be to increase the spectral resolution and the sensitivity of space missions. But, sooner or later, higher angular resolution will also be needed. Indeed, after the images of planets as single pixels, the imaging of planetary rings and moons will be the next step and later on, in a far but not unrealistic future, multipixel cartography of planet surfaces will be necessary to understand the exact ocean/continents morphology of planets, importance of polar caps, geographic repartition of {\it exo-vegetation} etc. 
\section{A Bottleneck in the Interferometric Approach}
Unfortunately, the first step of the interferometric approach in the Mid Infrared, i.e. Bracewell-like nulling interferometry with a few apertures, presents some difficult aspects: 
it requires high contrast/nulling performances  (like any other approach); 
the choosen nulling mechanism (Bracewell nulling) allows only a mono-pixel observation of the whole planetary system (with the subsequent problem of the background of exo-zodis); it requires high performance free flyers metrology; and by definition the sources (exoplanets) are extremely faint and contain only a few photons during the time scale of stability of the system.
By themselves, each of these difficulties are not insurmountable. The difficulty comes from their occurring at the same time.
One could imagine to start with a precursor. But the problem is that there seems to be no simple precursor that is 
technologically ``easy" and scientifically exciting.
\section{Breaking the Deadlock: a Two Step Solution}
Here we propose to decouple the needs for a thermal infrared approach and for future interferometers.
First a large IR coronagraph for close-by Super-Earths
and later on an interferometric precursor that is scientifically valuable and technically feasible.
Contrary to an interferometer with a few pupils, a large coronagraphic single aperture has an advantage:
it provides a full 2D image of the planetary systems instead of a single pixel signal, making the exo-zodi problem less severe. Other single aperture-like external occulters or Fresnel imagers are proposed, but they are better suited for the UV/Visible approach.
For the same angular resolution, the collecting area is at least 10 times larger than for the Darwin/TPF-I architecture.

The science requirements are the detection of H$_2$O, O$_3$ and CO$_2$ of the 70 nearest super-Earths (with a radius less that $2R_{\oplus}$) up to 5 pc.  They translate into a  20 m telescope (a deployable set a few 3-4 m
mirrors which would benefit of the JWST heritage) equipped with a coronograph and with a spectral resolution R = 20 to detect the central inversion peak of the CO2 band at 15 micron. With these characteristics the SNR would reach a value of 5 per spectral channel.
This type of architecture has already been studied  by the former TRW Company (presently Northrop Grumman) and submitted to JPL (Lillie et al. 2001).
The wavefront corrections and speckle calibration can be made e.g. by a ``Self-Coherent Camera" (Galicher et al 2008). They should be adaptable to the thermal infrared. METIS, the Mid-infrared ELT Imager and Spectrograph,  a proposed instrument for the European Extremely Large Telescope (E-ELT), is currently undergoing a phase-A study (Brandl et al 2008), and could be a ground-based precursor.

\section{A Scientifically Valuable Space Precursor}
Instead of accumulating problems due to free-flying control, faintness of exoplanets, single pixel detection, and high contrast, we propose to have a simpler precursor instrument which would still produce first class science:

- no nulling/high contrast

- bright targets

- no a priori wavelength constraints due to angular resolution (they are technically relaxed for free-flyers)

- different science: general astrophysics (not mainly exoplanets). For instance:
size and flattening of interesting stars,
astrometric perturbation of stellar centroid by transiting planets,
lensing (multiple quasars, « stellar arcs », astrometry of planetary lensing...),
super hot super-Earths.

The idea would be to start with 2 apertures, free-flying or not, 
interferometer with a evolution toward
a multi-aperture interferometer similar to the Stellar Imager (Carpenter et al. 2009).\\


{\bf References}

Brandl B., Lenzen R., Pantin E. et al., 2008. ETIS - the Mid-infrared E-ELT Imager and Spectrograph. SPIE 7014. astro-ph/0807.3271

Carpenter K., Schrijver C. and   Karovska M., 2009. The Stellar Imager (SI) project: a deep space UV/Optical Interferometer (UVOI) to observe the Universe at 0.1 milli-arcsec angular resolution. Planet. \& Spa. Sci. 320, 217

Galicher R., Baudoz P. and Rousset G., 2008. Wavefront error correction and Earth-like planet detection by Self-Coherent Camera in space, Astron. \& Astrophys., 488, L9

Lillie C. et al. 2001. 
TRW TPF Architecture. Phase 1 Study, Phase 2 Final Report. 
http://planetquest.jpl.nasa.gov/TPF/TPFrevue/FinlReps/Trw/TRW12Fnl.pdf


\end{document}